\documentclass[final,3p,times,twocolumn]{elsarticle}
\biboptions{comma,sort&compress}
\usepackage{here}
\usepackage{graphicx}
\usepackage{epsfig,epstopdf}
\usepackage{amsmath}
\usepackage{amssymb}
\usepackage{times}
\usepackage{wasysym}
\def\nin{\noindent}
\def\beq{\begin{equation}}
\def\eeq{\end{equation}}
\def\bea{\begin{eqnarray}}
\def\eea{\end{eqnarray}}

\journal{Nuc. Phys. (Proc. Suppl.)}

\begin{document}
\begin{frontmatter}

\title{The total top-pair production cross section at NNLL\tnoteref{preps}}
\tnotetext[preps]{TUM-HEP 853/12, TTK-12-36, ITP-UU-12/30, SPIN-12/28, FR-PHENO-2012-022, SFB/CPP-12-65}
 \author[label1,label3]{M. Beneke}
 
 \author[label2]{P. Falgari \corref{cor1}}
 \ead{p.falgari@uu.nl}
 \cortext[cor1]{Speaker}
 
 \author[label3]{S. Klein}
 \author[label1,label3]{J. Piclum}
 \author[label4]{C. Schwinn}
 \author[label3]{M. Ubiali}
 \author[label3]{F. Yan}
 
 \address[label1]{Physik Department T31,
James-Franck-Stra\ss e, Technische Universit\"at M\"unchen,
D--85748 Garching, Germany}
\address[label2]{Institute for Theoretical Physics and Spinoza Institute,
Utrecht University, 3508 TD Utrecht, The Netherlands}
\address[label3]{Institut f\"ur Theoretische Teilchenphysik und 
Kosmologie,
RWTH Aachen University, D--52056 Aachen, Germany}
\address[label4]{Albert-Ludwigs Universit\"at Freiburg, 
Physikalisches Institut, 
D-79104 Freiburg, Germany}

\begin{abstract}
\noindent
We present results for the total top-pair production cross section at the Tevatron and the LHC.
Our predictions supplement fixed-order results with resummation of soft logarithms and Coulomb 
singularities to next-to-next-to-leading (NNLL) logarithmic accuracy and include top-antitop bound-state effects.
The effects of resummation, the dependence on the PDF set used, the residual sources of theoretical uncertainty 
and their implication for measurements of the top-quark mass are discussed. 

\end{abstract}


\end{frontmatter}

\section{Introduction}
\nin
With hundreds of thousands of top quarks being produced yearly at 
the LHC, top-quark measurements are soon going to reach an unprecedented 
precision. In fact, the total top-pair production cross section has  
been measured by the two LHC collaborations with a total error of 
$\pm (4-6)\%$ \cite{ATLASxs, Collaboration:2012bt}, which is already comparable to, or below, 
the accuracy of analogous Tevatron results \cite{Abazov:2011cq,CDFxs}. 
With more statistics being collected at the LHC and a better understanding of 
systematic uncertainties, the error on the total $t \bar{t}$ production rate is 
bound to be reduced even further. 
Measurements of the inclusive cross section provide an important test of the 
Standard Model (SM) and constrain new-physics effects. They can also be
used to extract the top-quark mass in a theoretically clean way and to constrain 
the gluon PDF at medium-large values of the Bjorken variable $x$. Clearly, this 
is possible only if theoretical uncertainties are comparable to, or smaller than, 
the present experimental errors. 

The total theoretical uncertainty of a fixed-order next-to-leading order (NLO) 
calculation \cite{Nason:1987xz} is about $\pm 10\%$, bigger that the 
experimental accuracy at LHC and Tevatron. This motivates efforts to 
improve the available theoretical predictions. A full fixed-order NNLO calculation 
for the $q \bar{q}$ partonic production channel, which is 
relevant for Tevatron $t \bar{t}$ phenomenology, has been completed recently 
\cite{Baernreuther:2012ws,Czakon:2012zr}, while several ingredients, though not the full result yet,
are known for the $gg$ channel. 

In addition to a complete fixed-order NNLO calculation,
theoretical predictions can be improved by resumming sets of contributions 
known to all orders in perturbation theory. For the total cross section two classes 
of such corrections are relevant: threshold logarithms, which arise from soft-gluon 
emission, and Coulomb singularities, related to the potential interactions of the $t \bar{t}$ pair. 
Both corrections are enhanced near the partonic production threshold $\sqrt{\hat{s}}=2 m_t$, scaling respectively 
as $(\alpha_s \ln^{2,1} \beta)^n$ and $(\alpha_s/\beta)^n$, with the velocity $\beta$ of the final top (antitop) defined 
in terms of the partonic centre-of-mass energy as $\beta=\sqrt{1-4 m_t^2/\hat{s}}$.
Next-to-leading logarithmic (NLL) results for soft-log resummation have been available for a
while \cite{Kidonakis:1997gm,Bonciani:1998vc}, and recently next-to-next-to-leading logarithmic (NNLL) cross
sections resumming soft effects have been computed by several groups \cite{Kidonakis:2010dk, Aliev:2010zk,Ahrens:2011mw,Cacciari:2011hy},
thanks to a better understanding of the infrared structure of massive QCD amplitudes \cite{Becher:2009kw,Ferroglia:2009ep} and to the calculation of the
relevant anomalous dimensions \cite{Beneke:2009rj, Czakon:2009zw}. 
A combined resummation of soft and Coulomb corrections at NNLL accuracy, 
based on the soft-Coulomb factorization proven in \cite{Beneke:2010da}, 
has been presented in \cite{Beneke:2011mq}, and is the only
available prediction for the inclusive top-pair production cross section that resums both classes
of corrections, including effects from $t\bar{t}$ bound states below threshold. 
Explicit results of the calculation of \cite{Beneke:2011mq} are given in the following section.          
  
\section{The $t \bar{t}$ total cross section at NNLL}
\nin

The numerical results presented in this section are computed with the user-friendly program 
TOPIXS \cite{Beneke:2012wb}, which implements the NNLL soft-Coulomb resummation as 
described in \cite{Beneke:2011mq}. The resummed result for the $q\bar{q}$ channel is matched 
to the exact fixed-order NNLO cross section for this partonic channel \cite{Baernreuther:2012ws}, 
as detailed in Eq. (2.2) of  \cite{Beneke:2012wb}.  This gives (almost) full NNLO+NNLL accuracy 
at Tevatron, where the $q \bar{q}$ production channel dominates the hadronic cross section.   
For the matching of the $gg$ channel, which is dominant at the LHC, TOPIXS uses the approximated 
NNLO result of \cite{Beneke:2009ye}, which contains all the threshold-enhanced terms at NNLO,
but no constants at ${\cal O}(\alpha_s^4)$.
{\scriptsize 
\begin{table}[t]
\begin{center}{\small
\begin{tabular}{|l|c|c|c|}
\hline
$\sigma_{t \bar{t}}$[pb]&  Tevatron
& LHC (7 TeV) & LHC (8 TeV) \\
\hline
\hline
NLO & $6.68^{+0.36+0.23}_{-0.75-0.22}$ & $158.1^{+19.5+6.8}_{-21.2-6.2}$ & $226.2^{+27.8+9.2}_{-29.7-8.3}$ \\
\hline
NNLO & $7.00^{+0.21+0.29}_{-0.31-0.25}$ & $160.9^{+11.1+7.2}_{-11.5-6.7}$&  $229.8^{+16.5+9.7}_{-16.7-9.0}$ \\
\hline
{\bf NNLL} & $7.15^{+0.21+0.30}_{-0.20-0.25}$ & $162.4^{+6.7+7.3}_{-6.9-6.8}$&  $231.8^{+9.6+9.8}_{-9.9-9.1}$ \\
\hline
\end{tabular}}
\end{center}
\caption{Total $t \bar{t}$ cross section at NLO, NNLO and NNLL for Tevatron and LHC with $\sqrt{s}=7, \, 8\,$TeV and
$m_t=173.3\,$GeV. The first set of errors refers to the theoretical uncertainty, the second to the PDF+$\alpha_s$ uncertainty.
All the numbers are in picobarns.}
\label{tab:ttxs}
\end{table} }

In Table \ref{tab:ttxs} we present results for the total top-pair cross section at NLO, NNLO and 
(matched) NNLL accuracy for Tevatron and LHC with $\sqrt{s}=7,\, 8\,$GeV and $m_t=173.3\,$GeV
\footnote{In Table \ref{tab:ttxs} we use the same notation for Tevatron and LHC, though, strictly speaking, the 
LHC results are not exact at NNLO. The same is true for the matched NNLL cross section.}. The central
value for both renormalization and factorization scale is set to $m_t$. For the convolution of the partonic 
cross sections with the parton luminosities we use the MSTW2008 PDF sets \cite{Martin:2009iq} 
(NLO set for the NLO cross section, NNLO set for NNLO and NNLL cross sections). The two sets of errors refer to the theoretical uncertainty of the 
approximation and to the combined PDF and $\alpha_s$ error obtained with the $68\%$ confidence-level
PDF set. The theoretical uncertainty is obtained from scale variation for the NLO result, from the sum 
of scale uncertainty and ambiguities related to unknown ${\cal O}(\alpha_s^4)$ constant terms at NNLO
and from the sum of scale, constant and resummation uncertainties for the resummed NNLL result 
\cite{Beneke:2011mq}. Note that the error from the constant NNLO terms affects only the $gg$
channel, since the matching to the exact NNLO result for $q \bar{q}$ removes the uncertainty for
this channel.  
  
\begin{figure}[t!]
\begin{center}
\includegraphics[width=0.95 \linewidth]{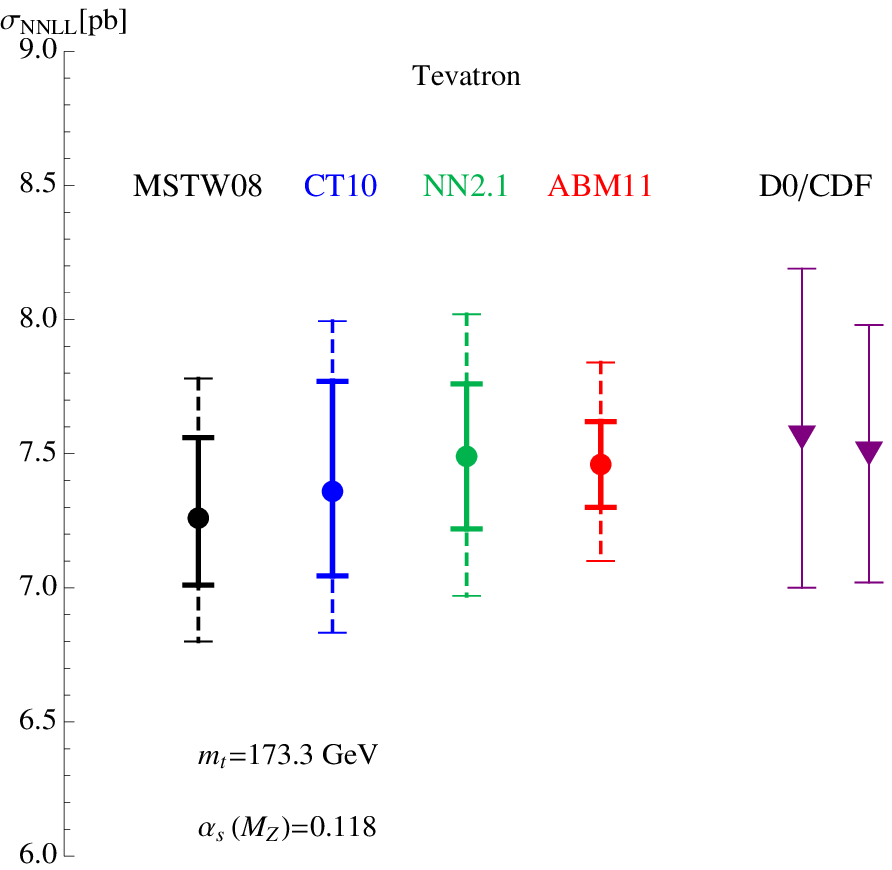}\\
\vspace{4 mm} 
\includegraphics[width=0.95 \linewidth]{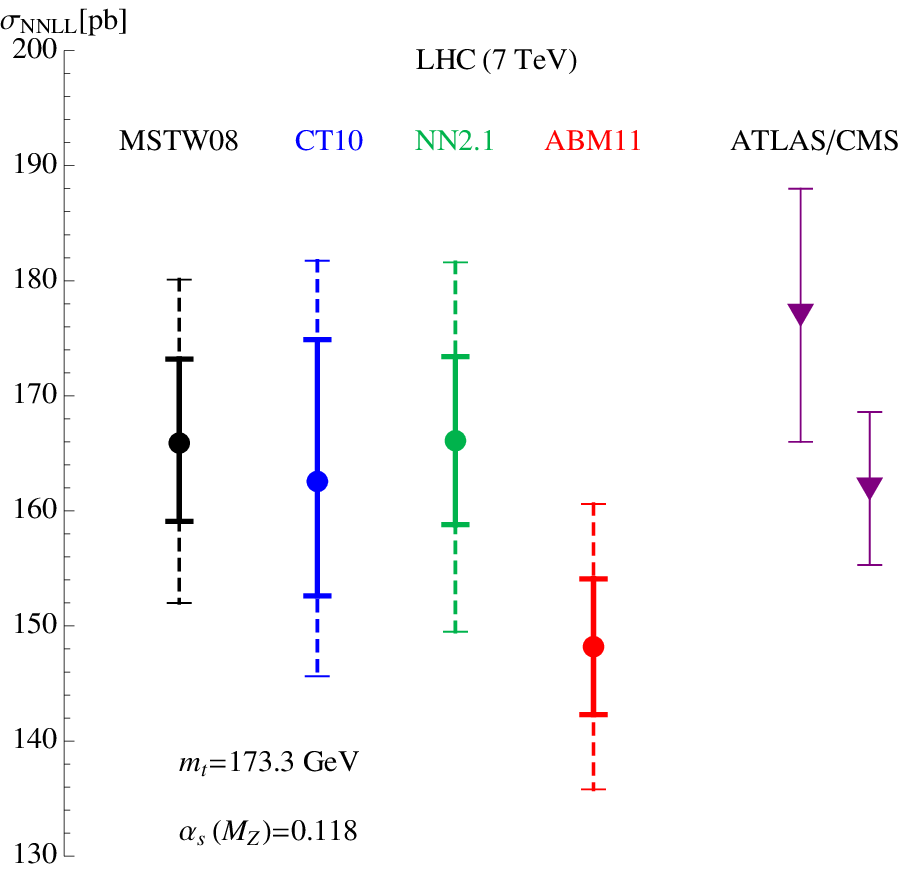}
\end{center}
\caption{Comparison of the NNLL predictions obtained with different PDF sets and of recent experimental measurements of the total
$t \bar{t}$ cross section, for Tevatron (upper plot) and LHC with $\sqrt{s}=7\,$TeV (lower plot). The two error bars for the theoretical 
numbers represent total theoretical uncertainty (external dashed bar) and the PDF+$\alpha_s$ uncertainty at $68\%$ confidence level. }
\label{fig:PDFs}
\end{figure}  
  
From Table \ref{tab:ttxs} it can be seen that at Tevatron corrections beyond NLO are sizeable, corresponding
to an upward shift of the cross section by $7 \%$. Of this, about $5\%$ is accounted for by NNLO contributions,
with higher-order terms from resummation contributing only $2 \%$ of the cross section. 
The situation is quite different at the LHC, where terms beyond NLO are only $3\%$ of the NLO cross section, of which 
only about $1\%$ originate from terms beyond ${\cal O}(\alpha_s^4)$. 
Note that the positive contribution of higher-order terms is partly compensated 
at the hadronic level by a downward shift due to
the switch from NLO to NNLO PDFs. While the effect of resummation is small for $t \bar{t}$ production, in \cite{Beneke:2012wb} it
has been shown that NNLL corrections beyond NLO can be much larger, up to $15-20\%$, for higher masses,
becoming relevant in the context of searches for new $\text{SU}(3)$-triplet fermions, e.g. in fourth generation searches.   

The addition of higher-order terms beyond NLO leads to a significant reduction of the theoretical uncertainty
on the $t \bar{t}$ cross section. This is particularly true at the Tevatron, where the inclusion of the exact 
NNLO result for the dominant $q \bar{q}$ channel removes completely the uncertainty
related to unknown ${\cal O}(\alpha_s^4)$ terms. The residual theory error for the NNLL result is about 
$\pm 3 \%$, smaller than the error of the most recent experimental measurements. At the LHC the remaining 
theoretical uncertainty is slightly larger ($\pm 4 \%$), due to the unknown NNLO constant contributions 
to the $gg$ production channel. Both at Tevatron and LHC the PDF+$\alpha_s$ error accounts for an additional 
$\pm (4-4.5) \%$ uncertainty.       

Since the PDF+$\alpha_s$ error represents now the largest individual source of uncertainty,
it is important to ascertain whether the error estimate provided by one of the many PDF parameterizations
available is consistent with the spread of the central-value predictions obtained with the different
sets. This is investigated in Figure \ref{fig:PDFs}, where the NNLL prediction for the total cross section 
obtained with the MSTW2008 \cite{Martin:2009iq}, CT10 \cite{Lai:2010vv}, NNPDF2.1 \cite{Ball:2011mu} and ABM11 \cite{Alekhin:2012ig} NNLO PDF sets 
are compared to each other and to the measurements provided by \cite{ATLASxs, Collaboration:2012bt, Abazov:2011cq, CDFxs}.
To make the comparison of the different sets more transparent we use a common 
initial value of the strong coupling constant, $\alpha_s(M_Z)=0.118$. At the Tevatron, the agreement
between different PDF sets is very good, and the central values for different PDF sets are compatible
with the error estimate of the individual sets. Furthermore, the NNLL theoretical predictions are 
remarkably close to the experimental values provided by D0 and CDF. 
At the LHC, MSTW2008, CT10 and NNPDF2.1 still show a good agreement with each other and 
with the experimentally measured cross sections. However, the prediction obtained with ABM11
differs significantly from the others, more than one would expect from the error estimate of the individual sets. 
This discrepancy is traceable to large differences in the gluon PDF between ABM11, which does not
include Tevatron jet data in its fits, and the other three sets in the region of medium-large Bjorken variable $x$, 
which is the most relevant to $t \bar{t}$ production. 

The difference observed at the LHC raises the question of whether measurements
of the $t \bar{t}$ cross section can be used to constrain the gluon PDF at medium-large $x$.     
This was investigated in \cite{Beneke:2012wb}, using the reweighting procedure of the NNPDF
collaboration to incorporate 
informations from recent $t \bar{t}$ measurements at the LHC and our NNLL prediction. It was found that 
the additional top-pair production input leads to a significant reduction of the uncertainties on the gluon parton luminosity. 
Also, in the case of the NNPDF2.1-DIS+DY sets, that do not include Tevatron jet data 
and whose gluon distribution is closer to the ABM11 one, the reweighting gives an upward shift of the 
gluon PDF that brings it close to the standard NNPDF2.1 distribution.  

\section{Top-mass extraction}
\nin
      
As pointed out in the introduction, measurements 
of the total $t \bar{t}$ cross section can be used to
extract the top-quark mass from data, 
as done, for example, in \cite{Abazov:2011pta} using different 
higher-order approximations for the cross section.  
Compared to a direct mass determination from 
the reconstruction of the top-quark decay products,
this method leads to larger uncertainties, but the 
extracted mass corresponds to a theoretically well-defined renormalization scheme, 
e.g $\overline{\text{MS}}$ or pole scheme.     
Here we show the effect of the inclusion of the exact
NNLO result for the $q \bar{q}$ channel and of higher-order
effects from NNLL resummation on the extraction of the 
pole mass $m_t$ from the Tevatron data.   
\begin{figure}
\begin{center}
\includegraphics[width=\linewidth]{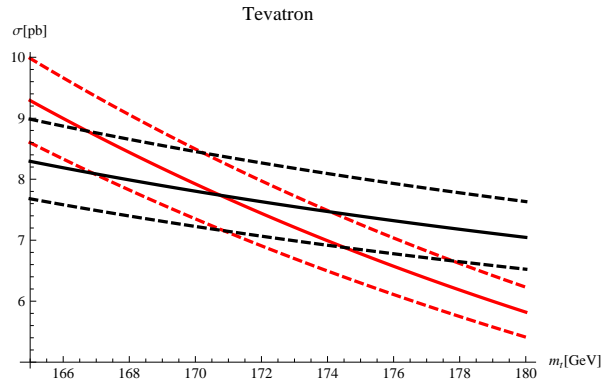}
\end{center}
\caption{Mass dependence of the experimental $t \bar{t}$ cross section  at Tevatron \cite{Abazov:2011cq} (black)
and of the NNLL resummed result provided by TOPIXS \cite{Beneke:2012wb} (red). Solid lines represent the 
central values, while dashed lines give the total experimental and theoretical uncertainties.}
\label{fig:xs_th_vs_exp}
\end{figure}

The central value of the pole mass $m_t$ is given by the maximum
of a likelihood function obtained from the convolution 
of two normalized gaussians centred at the theoretical
and experimental values respectively, with widths 
given by the total theoretical error, obtained from 
the linear sum of theory and PDF+$\alpha_s$ 
uncertainty, and experimental error. The mass 
dependence of the experimentally-measured 
cross section is obtained from \cite{Abazov:2011cq} and is plotted
in Figure \ref{fig:xs_th_vs_exp}, alongside the mass 
dependence of our NNLL result.      
 
Using our best NNLL result as theoretical 
input we extracted the pole mass 
$m_t = 171.4^{+5.4}_{-5.7}\,$GeV, 
in good agreement with the value from direct mass reconstruction
$m_t=173.2 \pm 0.8 \, \text{GeV}$. The value obtained using 
the approximated NNLO result as theory input is $m_t =  171.0^{+5.8}_{-6.3}$,
while the exact NNLO calculation gives 
$m_t =  170.5^{+5.7}_{-6.4}$. This shows that the effect of the 
exact ${\cal O}(\alpha_s^4)$ terms and higher-order contributions
from resummation on the central value is moderate, while a
reduction of the error is observed for the resummed result.

The CMS collaboration has recently published the most precise 
measurement of the $t \bar{t}$ cross section to date \cite{Collaboration:2012bt}, from which,
using our NNLL prediction, we obtained the pole mass 
$$m_t =  174.3^{+ 4.9}_{-4.4}\, \, \text{GeV}\,.$$
This is in even better agreement with the direct-reconstruction value 
and has an error of less than $\pm 3\%$. 
Note that all results shown are obtained assuming
that the Monte Carlo mass parameter which enters the determination
of the experimental cross section can be identified with the pole mass.
Allowing for a difference of $\pm 1 \,$GeV between the two masses
translates into an additional uncertainty of $\pm (0.4-0.5) \,$GeV 
on the extracted mass.  

\section{Conclusions}
\nin

We have presented updated results for the total top-pair 
cross section at Tevatron and LHC which 
include simultaneous resummation of soft and Coulomb 
effects, bound-state contributions and the recent exact NNLO 
result for the $q \bar{q}$ channel. Our best predictions,
\begin{eqnarray}
\text{Tevatron} &:& 7.15^{+0.21+0.30}_{-0.20-0.25}\, \text{pb}\, , \nonumber\\
\text{LHC} \,\, (\sqrt{s}=7\, \text{TeV}) &:& 162.4^{+6.7+7.3}_{-6.9-6.8}\, \text{pb}\, , \nonumber\\
\text{LHC} \,\, (\sqrt{s}=8\, \text{TeV}) &:& 231.8^{+9.6+9.8}_{-9.9-9.1}\, \text{pb}\, ,
\end{eqnarray}
show a good agreement with experimental measurements, and display a 
residual theoretical uncertainty of $\pm (3-4) \%$ and an additional
$\pm (4-4.5) \%$ error from the inputs for PDFs and $\alpha_s$. 
The dependence of the resummed result on different PDF sets
was found to be small at Tevatron, though a larger discrepancy 
between different PDF parameterizations is observed at the LHC.
Our NNLL prediction was used to extract the top-quark pole mass 
from Tevatron and LHC data, which resulted in values in good agreement 
with direct mass measurements and with a total error of the 
mass determination of $\pm 3 \%$ or better.



\end{document}